# Advancements in Nanoparticle-based Near-Infrared Fluorescence Probes for Cancer Specific Imaging Applications

Michael Adeniyi

Georgia Southern University/Department of Mechanical Engineering/Nanomaterial Research Lab, Statesboro GA

Email: adeniyitobi01@gmail.com

**Abstract**

Infrared (IR) dyes, especially those within the near-infrared (NIR) spectrum, offer substantial advantages for in vivo imaging, owing to their deep tissue penetration and minimal background autofluorescence. Nanoparticles incorporating these IR dyes, such as indocyanine green (ICG) have undergone extensive investigation regarding their pharmacokinetic behavior, including biodistribution patterns and clearance mechanisms. These nanoparticles often preferentially accumulate in tumor tissues, primarily through the enhanced permeability and retention (EPR) effect, thereby establishing them as potent instruments for sophisticated tumor imaging and precision-targeted drug delivery systems. The development of near-infrared fluorescence (NIRF) for large-scale tissue imaging is attributed to its ability to generate highly specific, targeted visualization of organs. The distinctive properties of the near-infrared spectrum, including its selective absorption by biomolecules that bind to specific sites within tissues, minimal autofluorescence, and reduced light scattering, make it an appealing option for cancer imaging. Unlike the conventional structural imaging system, molecular imaging leverages these properties to differentiate between malignant and normal tissues at the molecular level, offering a more refined and precise diagnostic capability. This review offers a timely and comprehensive overview of the latest advancements in Nanoparticle-based NIRF probes and multifunctional agents for cancer molecular imaging. These advances will extend the current concepts of cancer theranostics by NIRF imaging.

**Cancer Molecular Imaging**

Molecular imaging has introduced new possibilities for investigating and noninvasively tracking the formation, growth, and spread of tumors in living systems. It is anticipated to offer more detailed anatomical, physiological, and functional insights into diseases within clinical environments. Molecular imaging techniques hold significant potential as tools for the early detection of cancer, advancing drug discovery and development, and effectively tracking responses to treatments. Molecular imaging enables clinicians to identify the precise location of a tumor within the body and to observe the expression and activity of particular molecules (such as proteases and protein kinases), as well as biological processes (including apoptosis, angiogenesis, and metastasis) that impact tumor behavior and treatment response. This knowledge is anticipated to significantly influence cancer detection, personalized treatment strategies, drug development, and enhance our comprehension of cancer's origins [1]-[2]. There are several established imaging methods, such as positron emission tomography (PET), single photon emission computed tomography (SPECT), magnetic resonance imaging (MRI), optical bioluminescence, and optical fluorescence imaging, which are highly effective in capturing images of specific molecular targets associated with tumors [3]. Optical imaging techniques are becoming effective high-resolution methods for the early detection of cancer. They offer molecular insights into biological tissues and are linked to changes in important physiological parameters. Among the various optical imaging techniques, fluorescence imaging is particularly advantageous because it allows for the synchronous use of several different molecular probes. This technology operates using probes that emit within the near-infrared fluorescence (NIRF) spectrum, between 650 and 900 nm wavelengths. Polymethines, like Cy5.5 and Cy7, are the most commonly used organic NIR fluorophores for in-vivo NIRF imaging. Recently, several new NIRF probes characterized by improved fluorescence quantum yield, enhanced photostability, reduced aggregation, and lower cytotoxicity have been developed, including boron-dipyrromethene derivatives, Nile blue analogs, oxazine dye AOI987, and activatable probes. Advancements in NIRF-based in vivo cancer molecular imaging have accelerated significantly in recent years due to the creation of novel probes and improvements in optical imaging instruments [4].

In this review, we outline recent advancements in Nanoparticle-based near-infrared fluorescence (NIRF) probes for cancer molecular imaging. These nanoparticles, usually less than 200 nm in size, are designed to access nearly any part of the body. They can be easily modified with various targeting ligands and loaded with multiple contrast agent molecules significantly enhancing signal intensity across different imaging modalities. NIRF imaging utilizing nanoparticle-based probes is rapidly becoming a powerful tool in noninvasive cancer detection [5]. NIRF imaging offers multiple advantages over other optical imaging techniques for early cancer detection due to its low absorption and minimal autofluorescence. It also provides high sensitivity, the ability to detect multiple targets, and poses minimal risk to living subjects by using non-ionizing radiation. It is cost-effective, with affordable molecular probe preparation and easy-to-use detection equipment [6]. Also, the light signals released from biological tissues provide molecular information associated to pathophysiological changes and this technique relies on the use of a fluorescence probe that emits within the NIR spectrum [7].

There has been significant interest in the use of NIRF probes for both bioimaging and therapeutic applications. Their unique optical properties, combined with the ability to penetrate deep tissues and minimize background interference, make them highly promising tools for advancing diagnostics and targeted treatments in medical research. A wide range of NIRF dyes with excellent photophysical properties have been created. These dyes can be easily conjugated with various molecules, including small molecules, nucleotides, double-stranded DNA, DNA primers, amino acids, proteins, and antibodies, to achieve targeted specificity [8]. Their ability to be easily modified through conjugation with specific moieties makes NIRF dyes ideal for highly specific and sensitive cancer imaging. Nanoparticles encapsulating NIRF dyes and anticancer agents are crucial in the synergistic approach to cancer treatment [9].

**Classifications of organic NIRF dyes**

Despite the growing interest in NIRF dyes for bioimaging applications, their practical availability remains limited due to inherent challenges such as poor photostability, low hydrophilicity, and difficulties in effectively capturing signals within complex, heterogeneous tissues in-vivo. To address these issues, substantial research efforts have been devoted to reengineering and optimizing NIRF dyes. A significant focus has been placed on developing novel dyes with improved photostability and enhanced fluorescence intensity. Hydrophobic dyes have been chemically modified to increase hydrophilicity, thereby reducing self-aggregation and improving their performance in biological environments [10]. NIRF dyes are broadly categorized into several distinct groups based on their underlying organic fluorophore frameworks, including cyanines, rhodamine derivatives, 4,4-difluoro-4-bora-3a,4a-diaza-s-indacene (BODIPYs), squaraines, phthalocyanines, porphyrin analogs, and other structurally related dyes. Each class is defined by unique optical and chemical characteristics that enhance their suitability for precise near-infrared imaging across diverse biomedical settings [11]-[12]. The chemical structures of the dyes are shown in Figure 1.

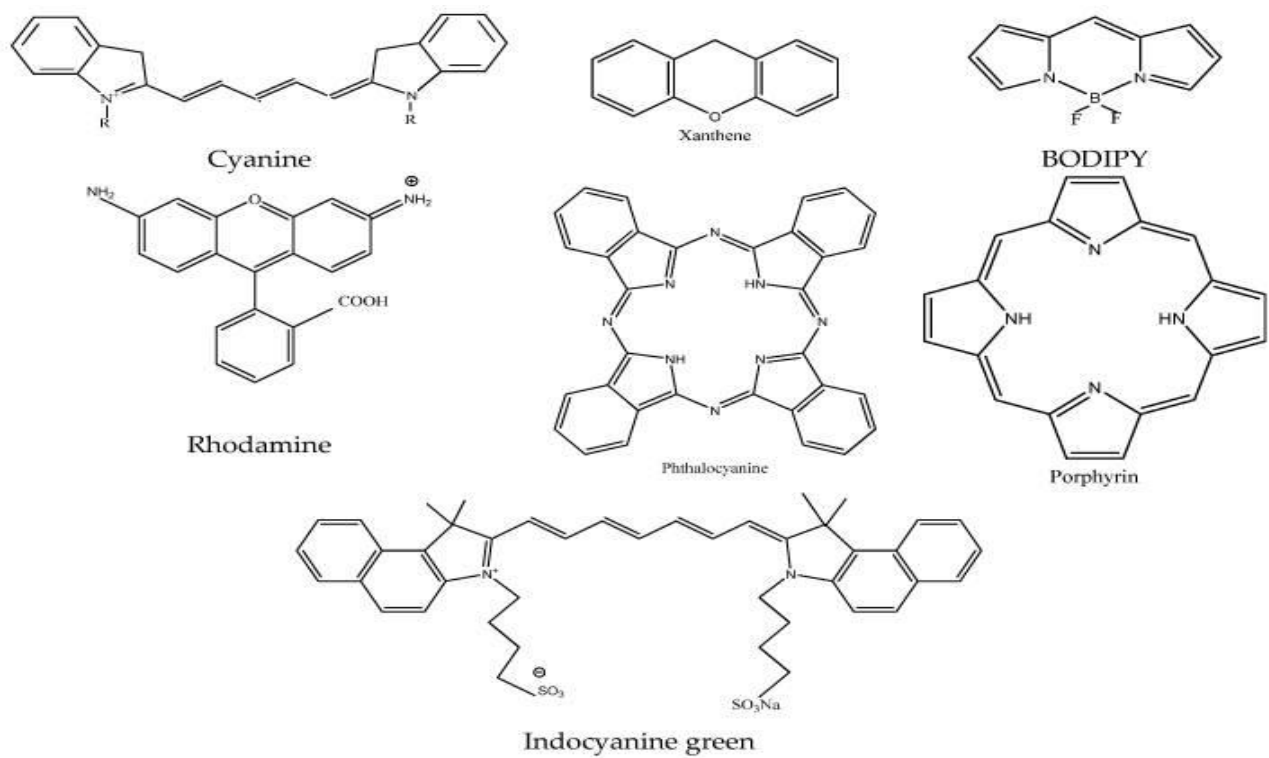

Figure 1. Basic chemical structure of near infrared fluorescent (NIRF) dyes.

Cyanine dyes

Cyanine dyes plays a key role in diagnostics and therapeutic research. Among these, indocyanine green (ICG)—a tricarbocyanine dye—was approved by the U.S. Food and Drug Administration (FDA) over five decades ago and remains widely used for clinical diagnostics. Structurally, cyanine dyes consist of two nitrogen-bearing heterocyclic rings functioning as charged chromophores, bridged by a polymethine chain with an odd number of carbons, which contributes to their distinct optical properties. Renowned for their narrow absorption spectra and high extinction coefficients, certain cyanine variants, such as Cy5 and Cy7, exhibit fluorescence within the near-infrared (NIR) range. This property makes them especially valuable as NIRF dyes, enabling high-intensity fluorescence and optimal tissue imaging due to their minimal background interference and strong signal clarity [4].

Rhodamine dyes

Rhodamine dyes, a prominent class within the xanthene dye family, have been extensively leveraged as fluorescent probes due to their superior photophysical properties, including high molar extinction coefficients and remarkable resistance to photobleaching. However, the conventional forms of these dyes emit fluorescence only within the visible light range (500–600 nm), restricting their applicability in in vivo bioimaging, where near-infrared (NIR) wavelengths are essential for deeper tissue penetration and reduced background interference. Significantly, the fluorescence behavior of rhodamine dyes can be finely tuned via various mechanisms, such as photo-induced electron transfer (PET) and reversible ring-opening/closing processes. By strategically modifying the xanthene core, researchers have synthesized NIR-emitting rhodamine derivatives. These modifications involve sophisticated molecular strategies, including PET modulation, redox reactions, and spirocyclic ring-opening mechanisms, allowing precise control over fluorescence intensity and wavelength. These advancements have transformed rhodamine dyes into highly versatile tools, expanding their applicability to NIR-based bioimaging and diagnostic platforms, which demand high signal clarity, photostability, and molecular specificity [9].

BODIPY-based NIRF probes

BODIPY (boron-dipyrromethene) established themselves as valuable tools for bioimaging, largely due to their high molar extinction coefficients, outstanding quantum yields, and exceptional thermal and photochemical stability. However, traditional BODIPY dyes are constrained by absorption and emission wavelengths that do not extend into the near-infrared (NIR) region, limiting their utility for deep tissue imaging where NIR fluorescence is crucial for minimizing autofluorescence and enhancing tissue penetration. To address these limitations, researchers have devised two principal approaches to transform BODIPY into NIR-active fluorophores. The first involves structural modification of the phenyl rings, while the second strategy merges the 3- and 5-phenyl rings with the aza-BODIPY core. These fusion forms six-membered rings, effectively reducing the torsional strain between the peripheral phenyl rings and the central core. By improving the co-planarity between these structural elements, these modifications facilitate a bathochromic shift, likely driven by enhanced electron delocalization across the conjugated system. This shift extends the optical activity into the NIR spectrum, enabling superior performance for in vivo bioimaging applications and fluorescence-based diagnostics. These advancements have significantly expanded the functional scope of BODIPY derivatives, making them indispensable in NIR fluorescence imaging for deep tissue visualization, molecular diagnostics, and theranostic platforms that demand stability, sensitivity, and high photonic output [9].

Squaraine-based NIRF probes

Squaraines, dyes, are recognized for their zwitterionic architecture, centered around an oxocyclobutenolate core. This core is flanked by aromatic or heterocyclic moieties, forming a donor–acceptor–donor structure, which contributes to their distinctive photophysical behavior. These dyes typically emit within the red to near-infrared (NIR) spectrum, offering desirable optical properties; however, their large molecular size and hydrophobicity present challenges for biological applications. To overcome these limitations, researchers have pursued three advanced strategies to enhance the performance of NIRF squaraine dyes, regulating aggregation and disaggregation, Functional Modifications with Targeting Moieties and Stimuli-Responsive Formation and Degradation. Recent innovations have further refined squaraine-based probes. For instance, bis(vinyl ruthenium)-modified squaraine dyes employ a reversible polyelectrochromic switch to modulate NIR absorption bands with precision, enhancing their adaptability in variable environments. Additionally, the integration of dicyanovinyl groups into the dye structure has significantly improved their NIR fluorescence intensity and chemical resilience, broadening their applicability in demanding biomedical settings such as deep tissue imaging and molecular diagnostics [6].

Phthalocyanines and porphyrin derivatives

Phthalocyanines and porphyrin derivatives represent a class of highly adaptable functional pigments, each comprising a core of four isoindole or pyrrole units with nitrogen atoms. Phthalocyanines are characterized by a $4n+2$ π-electron system, with the π-electrons extensively delocalized across the chromophore. This delocalization imparts exceptional chemical and thermal stability, allowing these compounds to withstand high-energy electromagnetic radiation. The two hydrogen atoms in the central cavity can be readily substituted with metal ions or functional groups, enabling precise modulation of their physicochemical properties to meet diverse application needs. While these pigments are utilized in fields such as electronics, optoelectronics, and biomedical science, their natural emission spectra fall short of the NIR range, limiting their utility in near-infrared fluorescence (NIRF) applications. To extend their emission into the NIR region, strategies such as introducing benzene ligands and incorporating electron-donating substituents have been employed. These modifications enhance π-conjugation and shift the emission towards NIR wavelengths. Porphyrins and their expanded analogs, which contain four or more pyrrole or heterocyclic rings, offer a robust framework for NIRF probe development. Their macrocyclic architecture provides not only excellent photophysical characteristics but also outstanding chemical stability, making them ideal candidates for advanced applications in molecular imaging, diagnostics, and targeted therapeutics. These structural advantages position porphyrin derivatives at the forefront of innovative NIRF probe design, enabling breakthroughs in fields such as bioimaging and theranostics [11].

**Nanoparticle-based NIRF probes**

NIRF nanoprobes are classified in two main categories: the well-established downconversion (DCN) NIRF nanoprobes and the newly developed upconversion (UCN) NIRF nanoprobes. DCN nanoprobes produce low-energy fluorescence when stimulated by high-energy light. Examples of these include nanoparticles embedded with NIRF dyes, quantum dots (QDs), single-walled carbon nanotubes (SWNTs), and metal nanoclusters. Conversely, UCN nanoprobes, a novel class of fluorescent nanoparticles, possess the capability to convert long-wavelength, low-energy excitation light into short-wavelength, high-energy fluorescence. A brief overview of the well-established NIRF nanoprobes including NIRF dyes containing nanoparticles, NIR fluorescent quantum dots (QDs), and gold nanoclusters are described below [13] [9].

NIRF dye-containing Nanoprobes

NIRF dye-containing nanoprobes consist of organic or inorganic matrix-based nanomaterials, where the NIRF dyes are either embedded within the matrix (dye-doped) or attached to the surface of the nanoparticles. For optimal performance, the matrix must be optically transparent to allow the excitation and emission light to pass through effectively [14]. Organic NIRF dyes are usually integrated via non-covalent or covalent bonding, creating functional nanoparticles with a core–shell structure. This design enhances photostability, biocompatibility, reduces self-aggregation, and delivers a bright fluorescence signal, with easily adjustable bioconjugation.

Gold nanostructures

Gold nanoparticles (AuNPs) are regarded as the top plasmonic material and have garnered attention in cancer research, particularly in cancer theranostics. AuNPs can be engineered into various shapes, including nanospheres, nanorods, nanoshells, nanostars, and nanocages, thanks to their exceptional tunability in size and shape as well as their favorable physiochemical properties. They have been widely studied for applications in Raman imaging and photoacoustic imaging (PAI), primarily because of their outstanding optical properties, high stability, low toxicity, and ease of bioconjugation. Fluorescent gold nanoprobes (GNPs), known for their ultra-small size, customizable optical properties, and high biocompatibility, serve as effective targeted probes for in vivo optical imaging [15].

Quantum dots

Quantum dots (QDs) are inorganic fluorescent semiconductor nanocrystals, they are synthesized in organic solvents like hexane, and consist of an inorganic core and shell, often metallic, which can be finely tuned to optimize peak fluorescence emission. Their use as fluorescent biotags in cell labeling and disease diagnostics stands out as a highly promising application in nanomedicine, owing to their distinctive optical properties. These include size-dependent photoluminescence, narrow and symmetric emission profiles, broad absorption spectra, and remarkable photostability. Such attributes position QDs as highly versatile tools for cutting-edge molecular imaging and diagnostic techniques in biomedical research. Quantum dot (QD) fluorescence emission can be precisely tuned to virtually any specific wavelength. More significantly, QDs exhibit exceptional resistance to photobleaching and chemical degradation, ensuring prolonged fluorescence stability. Additionally, their large surface area enables the conjugation of multiple targeting molecules to a single QD, enhancing their versatility in bioimaging applications. Furthermore, in organic solvents, QDs demonstrate extremely high quantum yields, making them highly efficient fluorescent probes for advanced imaging and diagnostic purposes [16].

**NIRF Probes designed with advanced targeting capabilities**

A critical challenge in the in vivo administration of NIR imaging agents lies in ensuring precise delivery and/or activation within the target tissue. To address this, a range of advanced strategies has been employed to enhance targeting specificity or probe activation, often intricately tailored to the molecular characteristics of the target. In the context of tumor targeting with NIRF nanoparticles, three principal methodologies have been adopted: (i) exploiting the enhanced permeability and retention (EPR) effect; (ii) molecular targeting via selective antigens or overexpressed receptors on cancer cell membranes; and (iii) chemical activation of nanoprobes in tumor-specific microenvironments, such as through enzymatic cleavage or oxidative processes [17]. Probes with targeting capabilities are pivotal for achieving high specificity and sensitivity in bioimaging applications. Integrating targeting moieties into these dyes or NIRF dye-loaded nanoparticles is a fundamental aspect of probe development. Presently, frequently employed targeting moieties include antibodies, peptides, proteins, aptamers, and small molecule receptor ligands. Remarkably, when these targeting elements are precisely integrated, the probes typically retain their intrinsic NIRF optical characteristics, along with their pharmacokinetic behavior and in vivo biodistribution [9].

NIRF dyes with enhanced tumor-targeting Capability

Numerous NIRF dyes have inherently demonstrated the capability to selectively accumulate in tumor tissues without the need for conjugation with targeting moieties. Notable examples include heptamethine indocyanine dyes such as IR-780 iodide, IR-783, MHI-148, and porphyrin derivatives like Pz 247, all of which exhibit preferential tumor localization. These properties confer distinct priority over conventional fluorescent agents like ICG. The multifunctional nature of these NIRF dyes provides deeper insights into tumor-specific imaging modalities, advancing the field of targeted diagnostics and therapeutic monitoring [18].

NIRF Nanoparticles with highly specific targeting capabilities

To achieve specific binding at targeted sites, NIRF dyes require conjugation with biomolecules that facilitate binding only in the targeted locations. A wide range of targeting groups have been developed to modify and functionalize NIRF nanoprobes, including small molecules, peptides, proteins, aptamers, engineered antibodies, and antibody-based ligands. As bioimaging technology advances, targeted NIRF probes are increasingly being used for cell labeling and tracking in physiological conditions, as well as for assessing the morphology and function of cancer cells and interstitial tissues [19]. These targeted NIRF nanoprobes have the potential to greatly improve delivery efficiency by increasing specificity and selectivity of targeting the desired tissue. Meanwhile, a range of molecules and proteins implicated in carcinogenesis and the tumor microenvironment have been integrated into the design of NIRF probes to confer highly specific targeting properties. [20].

Target-responsive NIRF nanoparticles

Activatable imaging probes enhance signal output in real time by reacting to specific biomolecular interactions or environmental changes [21]. Upon reaching the target site, enzyme cleavage triggers the release of fluorophores, which emit a strong fluorescent signal. This distinct feature of activatable fluorescence molecular probes offer significantly improved signal-to-background ratios and improved specificity for NIRF bioimaging, compared to traditional "constantly active" fluorescent probes. They also demonstrate excellent sensitivity and can penetrate deeply into tissues in vivo. Nanoparticle-based activatable probes have attracted considerable interest due to their NIRF quenching capabilities, which offer substantial advantages for high-precision optical imaging [22].

**Multimodality imaging**

While fluorescence serves as a robust technique for visualizing molecular targets at both microscopic and macroscopic levels, it does not inherently offer anatomical information—unless it is specifically directed towards structural markers such as the vasculature or cytoskeleton. Integrating NIR fluorescence imaging with complementary imaging modalities (e.g., optical, radionuclide, MRI, and CT) that provide anatomical context significantly enhances the value of the molecular data acquired through fluorescence. This multimodal approach facilitates more straightforward data interpretation, particularly in macroscopic imaging applications. These multimodal systems offer several unique advantages, such as enhancing the ability to quantify and precisely localize biological processes and events, facilitating the characterization of new imaging probes, and enabling the generation of near-perfectly aligned images with minimal or no motion artifacts. This, in turn, improves the interpretation and quantification of data across a wide range of experimental systems [23].

**Pharmacokinetics**

A thorough understanding of how NIRF nanoprobes enter, distribute within, and exit living organisms is crucial for designing nanoprobes optimized for molecular imaging. Various factors affect the probe's ability to reach its target, such as its binding affinity, the concentration of the target antigen, the excretion kinetics, the probe's physical dimensions, and the composition of the coating material, especially in nanoparticle-based formulations [17]. To achieve optimal accumulation of NIRF nanoprobes at tumor sites, it is essential to circumvent uptake by the reticuloendothelial system (RES) and the mononuclear phagocyte system (MPS). However, a significant challenge arises from the rapid clearance of many administered NIRF nanoprobes from the bloodstream, primarily due to RES and MPS activity, which leads to their sequestration in the liver and spleen. Consequently, the development of NIRF nanoprobes capable of evading swift systemic clearance represents a critical advancement for enhancing their functionality and precision as imaging agents. Three fundamental parameters that govern the in vivo behavior of NIRF nanoprobes include surface coating, particle size, and surface charge, they serve as key considerations for the strategic design of next-generation NIRF nanoprobes for cancer molecular imaging. Modifying the nanoprobe surface with hydrophilic polymers like polyethylene glycol (PEG) can improve solubility, reduce nonspecific interactions, extend systemic circulation, and improve tumor-specific targeting. However, such surface modifications may also result in a substantial increase in the overall nanoparticle size and the physical size of the imaging agent have a profound effect on its *in vivo* distribution [24]-[25]. Majority of NIRF nanoprobes employed for in vivo imaging are relatively large, with hydrodynamic diameters (HD) exceeding 20 nm, leading to preferential accumulation in the liver. From a clinical perspective, smaller nanoprobes are advantageous, as renal clearance is more efficient, thereby minimizing the risk of prolonged bodily retention. For instance, quantum dots with a neutral surface charge and a HD around 5.5 nm can undergo rapid renal elimination. However, the reduced size of these nanoprobes often correlates with shorter systemic half-lives, which typically compromises tumor accumulation. To optimize efficacy, NIRF nanoprobes should ideally be engineered to achieve a delicate balance: they must be small enough to enable efficient retention into tumor sites yet possess a sufficiently extended half-life to ensure adequate tumor targeting and retention [17].

**Toxicity**

Addressing and mitigating the potential in vivo toxicity of nanoprobes is paramount for their eventual transition to clinical application. Toxicity may stem either from the NIRF nanoprobes themselves or from degradation byproducts released in vivo. To address these challenges, various design strategies have been implemented to design biocompatible and biodegradable nanoprobes composed of minimally toxic constituents. For instance, clinically approved ICG fluorophores encapsulated within CPNPs exhibit both biocompatibility and biodegradability. Likewise, biodegradable NIR silicon nanoparticles have shown no detectable toxicity in preclinical animal models. Additionally, dye-loaded water-soluble polymers, designed for both biocompatibility and biodegradability, hold promise for clinical applications, offering a refined approach to minimizing toxicity while maintaining functionality in molecular imaging [26].

**Advancing Therapeutic Applications of Near-Infrared Fluorophores in Cancer Research**

Drug delivery

Chemotherapy has been widely applied to maximize the therapeutic outcome in cancer treatment. However, most cytotoxic drugs lack the ability of specific accumulation in tumors. In addition, various side effects may occur during the course of chemotherapy. These remain major impediments to the treatment of malignancies. Thus, novel platforms for targeted drug delivery that are safe and effective in vivo are highly desirable. Effective delivery of chemical drugs to tumor sites is particularly appealing for the enhancement of the tumor-killing effect and the reduction of systemic toxicities [27]. Studies have revealed that drug-loaded nanoparticle (NP) systems leverage the enhanced permeability and retention (EPR) effect to preferentially accumulate in tumor tissues, significantly enhancing drug bioavailability and extending therapeutic exposure. Furthermore, polymer conjugates demonstrate rapid absorption and sustained retention within the lymphatic system, while maintaining a favorable toxicity profile. At present, antibodies targeting specific cancer biomarkers and therapeutic targets have been successfully engineered. The strategic conjugation of monoclonal antibodies with NIRF dyes has further been employed for precision theranostics, facilitating targeted interventions in malignancies such as cutaneous tumors, breast, ovarian, gastric, and prostate cancers. This advanced, controlled drug delivery platform offers the potential to address the inherent limitations of conventional chemotherapy, particularly by improving tumor specificity and minimizing systemic side effects [3].

Photothermal therapy

Photothermal therapy (PTT), a noninvasive therapeutic modality, has garnered significant attention for its efficacy in treating various diseases, particularly in oncology. This advanced laser-induced thermal treatment, mediated by nanoparticles or pharmacological agents, represents one of the most promising strategies for inhibiting tumor growth with minimal morbidity and diminished systemic toxicity. The use of laser-absorbing agents or chromophores enhances localized thermal damage within the tumor microenvironment, amplifying the precision and effectiveness of the therapeutic intervention [28].

Photo-chemotherapy

Photo-chemotherapy utilizes light irradiation that is enhanced by photosensitizers to exert therapeutic effects in cancer tissues. When excited by light at a certain wavelength, photosensitizers facilitate the generation of cytotoxic free radicals [29]. These products affect tumor growth by destroying the abnormal neovasculature directly. They also initiate an inflammatory microenvironment that leads to cancer cell death. The first approved photosensitizer, Photofrin, is a composite of oligomeric porphyrins that has been applied for the treatment of lung cancer, esophagus cancer, and other types of cancer. The initial development of Photofrin in photo-chemotherapy for the treatment of bladder tumors was successful, current applications of Photofrin in photo-chemotherapy face numerous challenges that constrain its widespread clinical utility. These include insufficient penetration into deep tissues, poor tumor selectivity, and unintended accumulation in non-target areas such as the skin, leading to heightened photosensitivity upon exposure to sunlight. Therefore, the development of next-generation photosensitizers that circumvent these limitations is critical for the enhanced efficacy and broader application of photo-chemotherapy in oncology [30][31].

Near-Infrared Photoimmunotherapy

Near-Infrared Photoimmunotherapy (NIR-PIT) is based on cancer-targeted therapy that can selectively monitor and destroy cancer tissues. Nakajima et al developed a NIRF probe for PIT by linking a phthalocyanine dye IR700 and a monoclonal antibody. When exposed to NIR light, the conjugates that had been accumulated in the target sites induced highly specific tumoricidal activities. Selective binding avoided unnecessary injury to normal tissues. It was revealed that IR700 eventually accumulated in lysosomes. After exposure to a threshold intensity of NIR light, the conjugates immediately disrupted the outer cell membrane and lysosomes. Furthermore, repeated application of NIRF dyes was an effective strategy for cancer therapy without severe side effects; complete pathological remission might even be achieved [32].

**Conclusion**

Over the past decade, NIRF dyes have garnered substantial attention in biomedical research due to the NIR region's optimal optical window for deep tissue imaging. With appropriate modification, these dyes enable rapid cancer screening and early detection, serving as invaluable tools in guiding cancer therapies and contributing to reduced cancer morbidity and mortality rates. However, conventional dyes are susceptible to photobleaching and often fail to achieve adequate target-to-background contrast for clinical applications. Breakthroughs in nanotechnology and advanced imaging systems have catalyzed significant progress in probe design and multichannel imaging techniques. A wide array of nanoparticles has been engineered and evaluated as promising contrast agents or delivery vehicles for molecular imaging. The successful development of receptor-specific NIR nanoprobes, coupled with their applications in cancer diagnostics and therapeutic monitoring, has ignited substantial interest in personalized medical interventions, paving the way for more precise and individualized treatments. Nanoparticles (NPs) provide critical protection to NIRF dyes, safeguarding them from degradation while enhancing their fluorescence properties and enabling precise targeting with significantly improved signal-to-noise ratios. Despite the impressive outcomes observed in experimental research, the clinical translation of NP-based NIRF probes remains a distant prospect. Several substantial barriers need to be addressed before reaching the stage of clinical trials, particularly concerning the pharmacokinetic behavior and potential toxicity associated with the chemical and metallic materials used, optimal dosing strategies, and surface functionalization.